\documentclass[aps,prd,epsfig,preprint,amsmath,amssymb,superscriptaddress,showpacs,nofootinbib]{revtex4}
\usepackage{multirow}
\usepackage{graphicx,bm}
\usepackage[usenames]{color}
\usepackage{float}
\usepackage{caption}
\usepackage{subcaption}
\usepackage{slashed}
\begin{document}
\draft
\title{Model-independent study on the anomalous $ZZ\gamma$ and $Z\gamma\gamma$ couplings at the future muon collider}

\author{A. Senol}
\email[]{senol_a@ibu.edu.tr}
\affiliation{Department of Physics, Bolu Abant Izzet Baysal University, 14280, T\"{u}rkiye.}

\author{S. Spor}
\email[]{serdar.spor@beun.edu.tr}
\affiliation{Department of Medical Imaging Techniques, Zonguldak B\"{u}lent Ecevit University, 67100, Zonguldak, T\"{u}rkiye.}

\author{E. Gurkanli}
\email[]{egurkanli@sinop.edu.tr}
\affiliation{ Department of Physics, Sinop University, 57000, T\"{u}rkiye.}

\author{V. Cetinkaya}
\email[]{volkan.cetinkaya@dpu.edu.tr}
\affiliation{ Department of Physics, Kutahya Dumlupinar University, 43100, T\"{u}rkiye.}

\author{H. Denizli}
\email[]{denizli_h@ibu.edu.tr}
\affiliation{Department of Physics, Bolu Abant Izzet Baysal University, 14280, T\"{u}rkiye.}

\author{M. K{\"o}ksal}
\email[]{mkoksal@cumhuriyet.edu.tr}
\affiliation{Department of Physics, Sivas Cumhuriyet University, 58140, Sivas, T\"{u}rkiye.}

\date{\today}

\begin{abstract}
In this study, we investigate the potential of $\mu^{-} \mu^{+}\to Z\gamma \to \nu\bar{\nu}\gamma$ process at the future muon collider with a center-of-mass energy of 3 TeV to examine the anomalous $ZZ\gamma$ and $Z\gamma\gamma$ neutral triple gauge couplings defining $CP$-conserving $C_{\widetilde{B}W}/{\Lambda^4}$ coupling and three  $CP$-violating $C_{BB}/{\Lambda^4}$, $C_{BW}/{\Lambda^4}$, $C_{WW}/{\Lambda^4}$ couplings. All signal and relevant background events are generated in MadGraph and passed through Pythia for parton showering and hadronization. Detector effects are also considered via tuned muon detector cards in Delphes. The effects of systematic uncertainties of $0\%$, $3\%$ and $5\%$ on the sensitivities are studied. The best sensitivities obtained from the process $\mu^{-} \mu^{+}\to Z\gamma \to \nu\bar{\nu}\gamma$ are $[-6.53;6.64]\times10^{-2}$ TeV$^{-4}$ on $CP$-conserving $C_{\widetilde{B}W}/{\Lambda^4}$ coupling and $[-2.47;2.47]\times10^{-2}$ TeV$^{-4}$, $[-8.46;8.46]\times10^{-2}$ TeV$^{-4}$ and $[-2.20;2.20]\times10^{-1}$ TeV$^{-4}$ on $CP$-conserving  $C_{BB}/{\Lambda^4}$, $C_{BW}/{\Lambda^4}$, $C_{WW}/{\Lambda^4}$ couplings , respectively. Our obtained results on the anomalous neutral gauge couplings set more stringent sensitivity, ranging between 5 and 15 times than the current experimental results while slightly better than the phenomenological studies at future pp colliders such as the  HL-LHC,  the  HE-LHC and  the FCC-hh, respectively. On the other hand, we can see that the bounds on the anomalous neutral gauge couplings expected to be obtained for the future $e^{-}e^{+}$ colliders such as the CLIC are roughly 2 times better than our results.
\end{abstract}


\maketitle

\section{Introduction}

The trilinear and quartic gauge boson couplings defined by the non-Abelian $SU(2)_L \times U(1)_Y$ gauge theory of the Standard Model (SM) are of great interest since they provide valuable information to test the SM predictions at the TeV energy scale. The triple gauge boson couplings $WWV$, $ZV\gamma$ and $ZZV$(V=$\gamma$, $Z$) can be defined in the SM \cite{Baur:2000hfg}. Nevertheless, $Z\gamma\gamma$ and $ZZ\gamma$ couplings vanish at the tree-level in the SM due to the $Z$ boson having  no electrical charge. Therefore, possible deviations of the anomalous Neutral Triple Gauge Couplings (aNTGC) $Z\gamma\gamma$ and $ZZ\gamma$ from the SM predictions with the experimental results can give important clues about new physics coming from beyond the SM. For this purpose, in the literature, there have been many studies for aNTGC through the $ZZ$ and $Z\gamma$ production in the final state at $e^-e^+$ colliders \cite{Choudhury:1994ywq,Atag:2004ybz,Ots:2004twm,Ots:2006gsd,Rodriguez:2009rnw,Ananthanarayan:2012onz,Ananthanarayan:2014cal,Rahaman:2016nzs,Rahaman:2017qed,Ellis:2020ekm,Fu:2021jec,Ellis:2021rop,Yang:2022tgw,Spor:2022pou} and $pp$ colliders \cite{Baur:1993fkx,Senol:2018gvg,Rahaman:2019tnp,Senol:2019ybv,Senol:2020hbh,Yilmaz:2020ser,Yilmaz:2021dbm,Hernandez:2021wsz,Biekotter:2021ysx,Lombardi:2022plb,Hernandez-Juarez:2022kjx}. Vector boson pair production processes, which are also discussed in these studies, provide to test modifications of gauge-boson self-interactions, and the sensitivity is typically improved with increased collision energy. Although hadron colliders tend to provide the strongest limits on the aNTGC, a future multi-TeV lepton colliders would also be very competitive
\cite{Geng:2019ebo,Gounaris:2003lsd,Belloni:2022due}.

Muons and electrons are elementary particles in the SM, but the mass of the muon is about 207 times greater than that of the electron. This affects the size and cost of colliders and their achievable energies in particle collisions. The most important element in their discovery potential, as is known, is that particles must be accelerated to collide at the highest possible energies. But when the accelerator magnets bend the circular orbits of particles, they also lose energy through synchrotron radiation. The fact that protons and muons are heavier particles than electrons makes them lose much less energy. Therefore, although nowadays protons are used in the highest energy circular colliders, a small fraction of the energy from proton-proton collisions can be used to produce other particles, as collisions take place between quarks and gluons within the proton. At the same time, the use of composite particles in the proton-proton collider causes a lot of background, while the muon collider provides a clean environment as the background. In contrast to protons, all of the energy from their collisions can be utilized for particle generation \cite{Ryne:2020hye}. As a result, muons can investigate higher energy scales than protons colliding at the same beam energy \cite{Long:2021wja}. In other words, a muon collider with center-of mass energy of 14 TeV would have similar physics motivation and discovery potential to a proton-proton collider at the FCC-hh with center-of mass energy of 100 TeV \cite{Delahaye:2019egb}. However, on the way to a muon collider with a center-of-mass energy of multi-TeV, a 3 TeV machine stage appears as an attractive option. Emphasizing that muon colliders can be built gradually in strategic evaluations, it is suggested that the first stage with a 3 TeV muon collider can facilitate and accelerate the development of the project. The main advantage of the first stage is the significant reduction of the initial investment. In addition, possible delays are prevented by making concessions on technologies that have not yet been fully developed with the aim of more reasonable energy \cite{Blas:2022opk}. The ability to answer many open questions in particle physics with its prominent advantages increases the motivation of the planning studies on the future muon collider \cite{Palmer:2014asd,Antonelli:2016ezx,Wang:2016rwe,Neuffer:2018zxp,Boscolo:2019ytr,Bogomilov:2020twm}. In muon colliders, many possible phenomenological studies beyond the SM have recently made remarkable progress \cite{Buttazzo:2018wmg,Koksal:2019lja,Costantini:2020tkp,Yin:2020gre,Ruhdorfer:2020tgx,Chiesa:2020yhn,Bandyopadhyay:2021lja,Han:2021hrq,Liu:2021gtr,Han:2021twq,Capdevilla:2021xku,Bottaro:2021res,Capdevilla:2021ooc,Huang:2021edc,Asadi:2021wsd,Han:2021pas,Franceschini:2021pol,Chiesa:2021tyr,Buttazzo:2021eka,Huang:2022vke,Spor:2022kyz,Yang:2022dbn,Forslund:2022unz}.

A common way to probe new physics effects beyond the SM is by using an Effective Field Theory (EFT) approach. The way used to study the aNTGC with the SM gauge group consists of adding high-dimensional operators to the SM Lagrangian and obtaining the effective vertices running by the anomalous couplings \cite{Rahaman:2016nzs}.
 In this study, we focus on the dimension-eight operators defining the aNTGC. Therefore, the effective Lagrangian that contains SM interactions and new physics effects is defined by  \cite{Degrande:2014ydn}

\begin{eqnarray}
\label{eq.1}
{\cal L}^{\text{NTGC}}={\cal L}_{\text{SM}}+\sum_{i}\frac{C_i}{\Lambda^{4}}({\cal O}_i+{\cal O}_i^\dagger)
\end{eqnarray}

{\raggedright where $\Lambda$ shows the new physics scale, $C_i$ coefficients  are dimensionless parameters of the new physics and ${\cal O}_i$ dimension-eight operators  are given as follows}

\begin{eqnarray}
\label{eq.2}
{\cal O}_{\widetilde{B}W}=iH^{\dagger} \widetilde{B}_{\mu\nu}W^{\mu\rho} \{D_\rho,D^\nu \}H,
\end{eqnarray}
\begin{eqnarray}
\label{eq.3}
{\cal O}_{BW}=iH^\dagger B_{\mu\nu}W^{\mu\rho} \{D_\rho,D^\nu \}H,
\end{eqnarray}
\begin{eqnarray}
\label{eq.4}
{\cal O}_{WW}=iH^\dagger W_{\mu\nu}W^{\mu\rho} \{D_\rho,D^\nu \}H,
\end{eqnarray}
\begin{eqnarray}
\label{eq.5}
{\cal O}_{BB}=iH^\dagger B_{\mu\nu}B^{\mu\rho} \{D_\rho,D^\nu \}H
\end{eqnarray}

{\raggedright where}

\begin{eqnarray}
\label{eq.6}
B_{\mu\nu}=\left(\partial_\mu B_\nu - \partial_\nu B_\mu\right),
\end{eqnarray}
\begin{eqnarray}
\label{eq.7}
W_{\mu\nu}=\sigma^i\left(\partial_\mu W_\nu^i - \partial_\nu W_\mu^i + g\epsilon_{ijk}W_\mu^j W_\nu^k\right),
\end{eqnarray}

{\raggedright with $\langle \sigma^i\sigma^j\rangle=\delta^{ij}/2$ and}

\begin{eqnarray}
\label{eq.8}
D_\mu \equiv \partial_\mu - i\frac{g^\prime}{2}B_\mu Y - ig_W W_\mu^i\sigma^i.
\end{eqnarray}

Given in the definitions above, $B_{\mu \nu}$ and $W_{\mu \nu}$ are the field strength tensors, $H$ is the Higgs field  and $D_{\mu}$ is the covariant derivative. 
{\raggedright Here, while ${\cal O}_{\widetilde{B}W}$ operator is $CP$-conserving, ${\cal O}_{BW}$, ${\cal O}_{WW}$ and ${\cal O}_{BB}$ operators are $CP$-violating.}

The interference between the SM and the dimension-eight operators provides the largest new physics contribution for $Z\gamma$ production when the energy scale is high. Unless interference between the SM and the dimension-eight and dimension-ten operators are both powerfully suppressed, the ${\cal O}({\Lambda^{-8}})$ square of dimension-eight operators cannot be expected to include a contribution from the new physics at a high-energy scale. At the tree-level, the dimension-six operators have no effect on aNTGC while the contributions of the dimension-eight operators are of the order ${\upsilon^2\hat{s}}/{\Lambda^4}$. However, the dimension-six operators induce an effect with the order ${\alpha \hat{s}}/{4\pi\Lambda^2}$ on aNTGC at the one-loop. Accordingly, the one-loop contribution of the dimension-six operators is negligible compared to the contribution of the dimension-eight operators for $\Lambda \lesssim \sqrt{4\pi\hat{s}/\alpha}$ \cite{Degrande:2014ydn}. Here, $\alpha$ and $\upsilon$ are fine-structure constant and vacuum expectation value, respectively. 

On the other hand, effective Lagrangian for the aNTGC with other dimension-six and dimension-eight operators examined in the literature is generally described as follows \cite{Gounaris:2000svs}

\begin{eqnarray}
\label{eq.9}
\begin{split}
{\cal L}_{\text{aNTGC}}^{\text{dim-six,eight}}=&\frac{e}{m_Z^2}\Bigg[-[f_4^\gamma(\partial_\mu F^{\mu\beta})+f_4^Z(\partial_\mu Z^{\mu\beta})]Z_\alpha (\partial^\alpha Z_\beta)+[f_5^\gamma(\partial^\sigma F_{\sigma\mu})+f_5^Z (\partial^\sigma Z_{\sigma\mu})]\widetilde{Z}^{\mu\beta}Z_\beta  \\
&-[h_1^\gamma (\partial^\sigma F_{\sigma\mu})+h_1^Z (\partial^\sigma Z_{\sigma\mu})]Z_\beta F^{\mu\beta}-[h_3^\gamma(\partial_\sigma F^{\sigma\rho})+h_3^Z(\partial_\sigma Z^{\sigma\rho})]Z^\alpha \widetilde{F}_{\rho\alpha}   \\
&-\bigg\{\frac{h_2^\gamma}{m_Z^2}[\partial_\alpha \partial_\beta \partial^\rho F_{\rho\mu}]+\frac{h_2^Z}{m_Z^2}[\partial_\alpha \partial_\beta(\square+m_Z^2)Z_\mu]\bigg\}Z^\alpha F^{\mu\beta}   \\
&+\bigg\{\frac{h_4^\gamma}{2m_Z^2}[\square\partial^\sigma F^{\rho\alpha}]+\frac{h_4^Z}{2m_Z^2}[(\square+m_Z^2)\partial^\sigma Z^{\rho\alpha}]\bigg\}Z_\sigma\widetilde{F}_{\rho\alpha}\Bigg].
\end{split}
\end{eqnarray}

Here, $\widetilde{Z}_{\mu\nu}=1/2\epsilon_{\mu\nu\rho\sigma}Z^{\rho\sigma}$ $(\epsilon^{0123}=+1)$ with field strength tensor $Z_{\mu\nu}=\partial_\mu Z_\nu - \partial_\nu Z_\mu$ and for the electromagnetic field tensor $F_{\mu\nu}$. However, $f_4^V$, $h_1^V$, $h_2^V$ are the $CP$-violating couplings while $f_5^V$, $h_3^V$, $h_4^V$ are the $CP$-conserving couplings $(V=\gamma$, $Z)$. In the SM, all these couplings at tree-level are zero. In the Lagrangian, $h_2^V$ and $h_4^V$ couplings are related to dimension-eight while the remaining four couplings are related to dimension-six.

When $SU(2)_L \times U(1)_Y$ gauge invariance is taken into account, the couplings in Eq.~(\ref{eq.9}) are correlated with the couplings of the operators given in Eqs.~(\ref{eq.2}-\ref{eq.5}) \cite{Rahaman:2020fdf}. The $CP$-conserving anomalous couplings with the two on-shell $Z$ bosons in addition to one off-shell $V=\gamma$ or $Z$ boson are given by \cite{Degrande:2014ydn}

\begin{eqnarray}
\label{eq.10}
f_5^Z=0,
\end{eqnarray}
\begin{eqnarray}
\label{eq.11}
f_5^\gamma=\frac{\upsilon^2 m_Z^2}{4c_\omega s_\omega} \frac{C_{\widetilde{B}W}}{\Lambda^4}.
\end{eqnarray}

{\raggedright However, the $CP$-violating anomalous couplings are given as}

\begin{eqnarray}
\label{eq.12}
f_4^Z=\frac{m_Z^2 \upsilon^2 \left(c_\omega^2 \frac{C_{WW}}{\Lambda^4}+2c_\omega s_\omega \frac{C_{BW}}{\Lambda^4}+4s_\omega^2 \frac{C_{BB}}{\Lambda^4}\right)}{2c_\omega s_\omega},
\end{eqnarray}
\begin{eqnarray}
\label{eq.13}
f_4^\gamma=-\frac{m_Z^2 \upsilon^2 \left(-c_\omega s_\omega \frac{C_{WW}}{\Lambda^4}+\frac{C_{BW}}{\Lambda^4}(c_\omega^2-s_\omega^2)+4c_\omega s_\omega \frac{C_{BB}}{\Lambda^4}\right)}{4c_\omega s_\omega}.
\end{eqnarray}

Here, $c_\omega$ and $s_\omega$ are the cosine and sine of weak mixing angles $\theta_{w}$. On the other hand, the $CP$-conserving anomalous couplings with one on-shell $Z$ boson and photon in addition to one off-shell $V=\gamma$ or $Z$ boson are written as follows \cite{Degrande:2014ydn}

\begin{eqnarray}
\label{eq.14}
h_3^Z=\frac{\upsilon^2 m_Z^2}{4c_\omega s_\omega} \frac{C_{\widetilde{B}W}}{\Lambda^4},
\end{eqnarray}
\begin{eqnarray}
\label{eq.15}
h_4^Z=h_3^\gamma=h_4^\gamma=0.
\end{eqnarray}

{\raggedright Also, the $CP$-violating anomalous couplings are described by the following formulas}

\begin{eqnarray}
\label{eq.16}
h_1^Z=\frac{m_Z^2 \upsilon^2 \left(-c_\omega s_\omega \frac{C_{WW}}{\Lambda^4}+\frac{C_{BW}}{\Lambda^4}(c_\omega^2-s_\omega^2)+4c_\omega s_\omega \frac{C_{BB}}{\Lambda^4}\right)}{4c_\omega s_\omega},
\end{eqnarray}
\begin{eqnarray}
\label{eq.17}
h_2^Z=h_2^\gamma=0,
\end{eqnarray}
\begin{eqnarray}
\label{eq.18}
h_1^\gamma=-\frac{m_Z^2 \upsilon^2 \left(s_\omega^2 \frac{C_{WW}}{\Lambda^4}-2c_\omega s_\omega \frac{C_{BW}}{\Lambda^4}+4c_\omega^2 \frac{C_{BB}}{\Lambda^4}\right)}{4c_\omega s_\omega}.
\end{eqnarray}

The dimension-eight coefficients given in Eqs.~(\ref{eq.11}-\ref{eq.14},\ref{eq.16},\ref{eq.18}) are described aNTGC as $CP$-conserving $C_{\widetilde{B}W}/{\Lambda^4}$ and $CP$-violating $C_{BB}/{\Lambda^4}$, $C_{BW}/{\Lambda^4}$, $C_{WW}/{\Lambda^4}$. 

The best experimental limits obtained at 95$\%$ Confidence Level (C.L.) on dimension-eight couplings $C_{\widetilde{B}W}/{\Lambda^4}$, $C_{BB}/{\Lambda^4}$, $C_{BW}/{\Lambda^4}$ and $C_{WW}/{\Lambda^4}$ are investigated via process $pp\rightarrow Z\gamma \rightarrow \nu\bar{\nu}\gamma$ at $\sqrt{s}=13$ TeV with $L_{int}=36.1$ fb$^{-1}$ at the LHC \cite{Aaboud:2018ybz}. These are given as

\begin{eqnarray}
\label{eq.19}
-1.1\, \text{TeV}^{-4}<\frac{C_{\widetilde{B}W}}{\Lambda^4}<1.1 \, \text{TeV}^{-4},
\end{eqnarray}
\begin{eqnarray}
\label{eq.20}
-2.3\, \text{TeV}^{-4}<\frac{C_{WW}}{\Lambda^4}<2.3 \, \text{TeV}^{-4},
\end{eqnarray}
\begin{eqnarray}
\label{eq.21}
-0.65\, \text{TeV}^{-4}<\frac{C_{BW}}{\Lambda^4}<0.64 \, \text{TeV}^{-4},
\end{eqnarray}
\begin{eqnarray}
\label{eq.22}
-0.24\, \text{TeV}^{-4}<\frac{C_{BB}}{\Lambda^4}<0.24 \, \text{TeV}^{-4}.
\end{eqnarray}

Phenomenological studies on the sensitivities to $CP$-conserving $C_{\widetilde{B}W}/{\Lambda^4}$ and $CP$-violating $C_{BB}/{\Lambda^4}$, $C_{BW}/{\Lambda^4}$ and $C_{WW}/{\Lambda^4}$ couplings have been performed through the process $pp\,\rightarrow\,ZZ\,\rightarrow\,4\ell$ at FCC-hh \cite{Yilmaz:2020ser}, the process $pp\,\rightarrow\,Z\gamma\,\rightarrow\,\nu\nu\gamma$ at HL-LHC and HE-LHC \cite{Senol:2020hbh} and the process $e^-e^+\,\rightarrow\,Z\gamma\,\rightarrow\,\nu\nu\gamma$ at CLIC \cite{Spor:2022pou}. The limits of anomalous $C_{\widetilde{B}W}/{\Lambda^4}$, $C_{BB}/{\Lambda^4}$, $C_{BW}/{\Lambda^4}$ and $C_{WW}/{\Lambda^4}$ couplings at 95$\%$ C.L. have been found as $[-11.7;11.7]\times10^{-2}$ TeV$^{-4}$, $[-13.8;13.8]\times10^{-2}$ TeV$^{-4}$, $[-38.0;37.9]\times10^{-2}$ TeV$^{-4}$ and $[-2.93;2.92]\times10^{-1}$ TeV$^{-4}$ at the FCC-hh collider, $[-38.0;38.0]\times10^{-2}$ TeV$^{-4}$, $[-21.0;21.0]\times10^{-2}$ TeV$^{-4}$, $[-48.0;48.0]\times10^{-2}$ TeV$^{-4}$ and $[-10.8;10.8]\times10^{-1}$ TeV$^{-4}$ at the HL-LHC collider, $[-12.0;12.0]\times10^{-2}$ TeV$^{-4}$, $[-8.5;8.5]\times10^{-2}$ TeV$^{-4}$, $[-25.0;25.0]\times10^{-2}$ TeV$^{-4}$ and $[-3.8;3.8]\times10^{-1}$ TeV$^{-4}$ at the HE-LHC collider and $[-4.28;4.52]\times10^{-2}$ TeV$^{-4}$, $[-1.35;2.00]\times10^{-2}$ TeV$^{-4}$, $[-4.75;5.24]\times10^{-2}$ TeV$^{-4}$ and $[-1.17;1.12]\times10^{-1}$ TeV$^{-4}$ at the CLIC collider, respectively.

The minimum coupling value of the coefficients is necessary, according to the EFT approach, to put the operator scale $\Lambda$ beyond the kinematic range of the distributions. The coefficients of dimension-eight operators can be connected to the new physics characteristic scale $\Lambda$, and an upper bound can be imposed on this scale \cite{Senol:2020hbh}. For $C=O(1)$ couplings, we calculate $\Lambda<\sqrt{4\pi \upsilon \sqrt{s}}\sim3.04$ TeV. For this reason, we focus on the process $\mu^-\mu^+\,\rightarrow\,Z\gamma \rightarrow \nu\bar{\nu}\gamma$ to examine for effects of the dimension-eight anomalous $ZZ\gamma$ and $Z\gamma\gamma$ couplings at the proposed future muon collider with an only center-of-mass energy of 3 TeV.

\section{Generation of signal and background events} \label{Sec2}

In this study, we investigate $C_{\widetilde{B}W}/{\Lambda^4}$, $C_{BB}/{\Lambda^4}$, $C_{BW}/{\Lambda^4}$ and $C_{WW}/{\Lambda^4}$ couplings defining the anomalous $ZZ\gamma$ and $Z\gamma\gamma$ couplings through the process $\mu^-\mu^+\,\rightarrow\,Z\gamma \rightarrow \nu\bar{\nu}\gamma$ at the muon collider with $\sqrt{s}$=3 TeV and $L_{int}$=1 ab$^{-1}$. We produce signal and background events analyses for the process $\mu^-\mu^+\,\rightarrow\,Z\gamma \rightarrow \nu\bar{\nu}\gamma$ by importing the signal aNTGC implemented via the UFO model file into {\sc MadGraph5}$\_$aMC@NLO \cite{Alwall:2014cvc}. Also, for parton showering and hadronization, the PYTHIA 8.2 package is used \cite{Sjostrand:2015rve}. $600$k events of the signal and the relevant backgrounds for each anomalous coupling are generated. Using the description of the muon collider detector card  implemented in Delphes 3.4.1 the detector response is simulated \cite{Favereau:2014qaz,muoncard}. Signal and background events are investigated by using the ExRootAnalysis \cite{ExRootAnalysis} package with ROOT 6.16 \cite{Brun:1997cvb}.

We examine $\nu\bar{\nu}\gamma$ final state assuming the decay of the Z boson into neutrinos in the $Z\gamma$ production. Our process involving $\nu\bar{\nu}\gamma$ in the final state has some advantages concerning  processes involving $q\bar{q}\gamma$ (hadronic channel of the Z boson) or $\ell^{+}\ell^{-}\gamma$ (leptonic channel of the Z boson) productions.  Here, $q\bar{q}\gamma$ channel does not have clean data due to a large number of QCD backgrounds. Also, as is known, $\nu\bar{\nu}\gamma$ channel has a higher branching ratio than $\ell^{+}\ell^{-}\gamma$ channel. Therefore, $\nu\bar{\nu}\gamma$ channel enables the opportunity to investigate $Z\gamma$ production in the more energetic region where the sensitivity is high \cite{Aaboud:2018ybz}.

\begin{figure}[ht]
\centerline{\scalebox{0.9}{\includegraphics{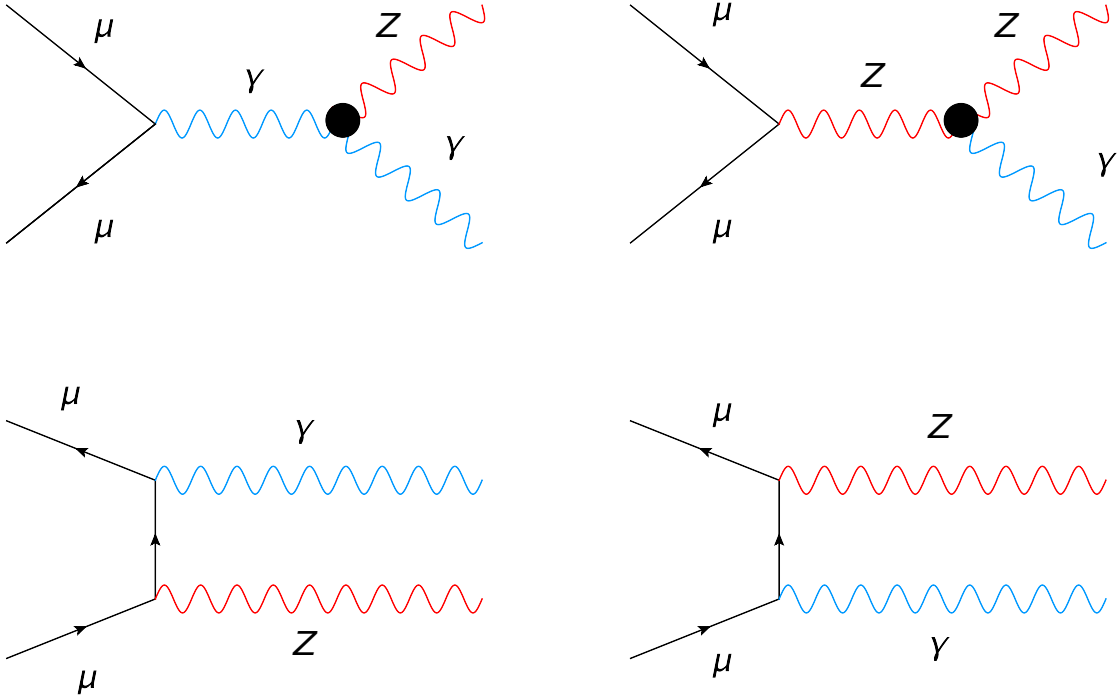}}}
\caption{Feynman diagrams of the process $\mu^- \mu^{+} \to Z\gamma $ with the anomalous $ZZ\gamma$ and $Z\gamma\gamma$ vertices(top) and the SM diagrams(bottom).}
\label{Fig.1}
\end{figure}

In Fig.~\ref{Fig.1}, the Feynman diagrams for the process $\mu^-\mu^+\,\rightarrow\,Z\gamma$ are represented. Here, the first two Feynman diagrams include contributions of the signal arising from the anomalous $ZZ\gamma$ and $Z\gamma\gamma$ couplings, while the other diagrams show the SM contributions. Since we examine the aNTGC through the process $\mu^-\mu^+\,\rightarrow\,Z\gamma \rightarrow\, \nu\bar{\nu}\gamma$ at the muon collider, a photon and missing energy constitute the final state topology of the signal process. In this case, the relevant SM background processes with the same or similar final state topology are taken into account as follows

\begin{eqnarray}
\label{eq.23}
\mu^-\mu^+\,\rightarrow\,Z\gamma \rightarrow\, \nu\bar{\nu}\gamma \,\,(SM),
\end{eqnarray}

\begin{eqnarray}
\label{eq.24}
\mu^-\mu^+\,\rightarrow\,\gamma \gamma \,\, (\gamma \gamma),
\end{eqnarray}

\begin{eqnarray}
\label{eq.25}
\mu^-\mu^+\,\rightarrow\,Z \gamma \rightarrow\, \ell\ell\gamma \,\, (\ell\ell\gamma),
\end{eqnarray}

\begin{eqnarray}
\label{eq.26}
\mu^-\mu^+\,\rightarrow\,WW\gamma \,\,(WW\gamma).
\end{eqnarray}

\begin{figure}[ht]
\centerline{\scalebox{0.60}{\includegraphics{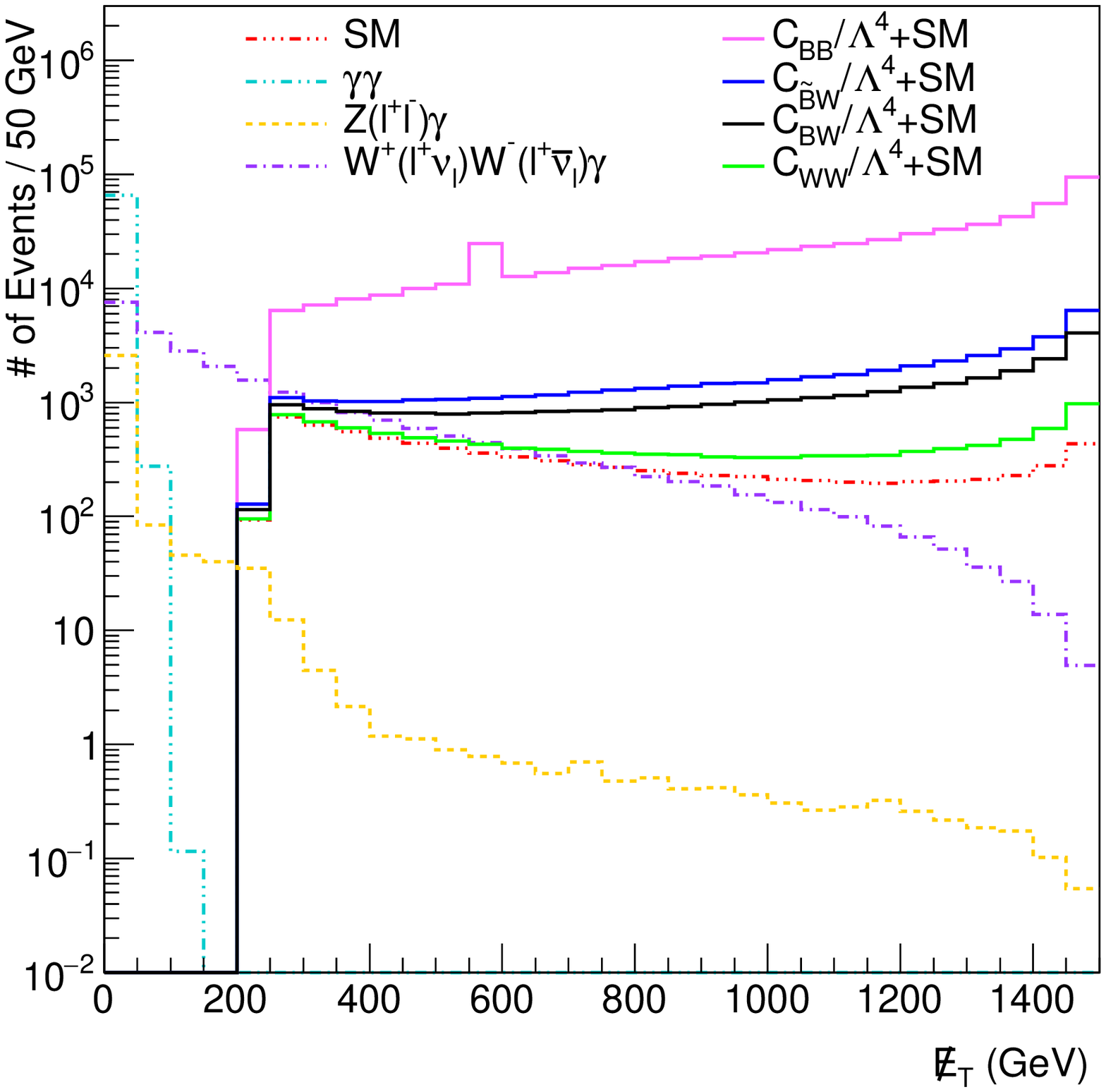}}}
\caption{The number of expected events (normalized to the $L_{int}$=1 ab$^{-1}$) as a function of $\slashed{E}_T$ missing energy transverse for the $\mu^+\mu^- \to Z\gamma$ signal and backgrounds. Here, $C_{\widetilde{B}W}/{\Lambda^4}$, $C_{BB}/{\Lambda^4}$, $C_{BW}/{\Lambda^4}$ and $C_{WW}/{\Lambda^4}$ couplings are equal to 1 TeV$^{-4}$. Here, the dotted lines represent the background processes, while the solid lines indicate the signal processes.}
\label{Fig.2}
\end{figure}

\begin{figure}[ht]
\centerline{\scalebox{0.60}{\includegraphics{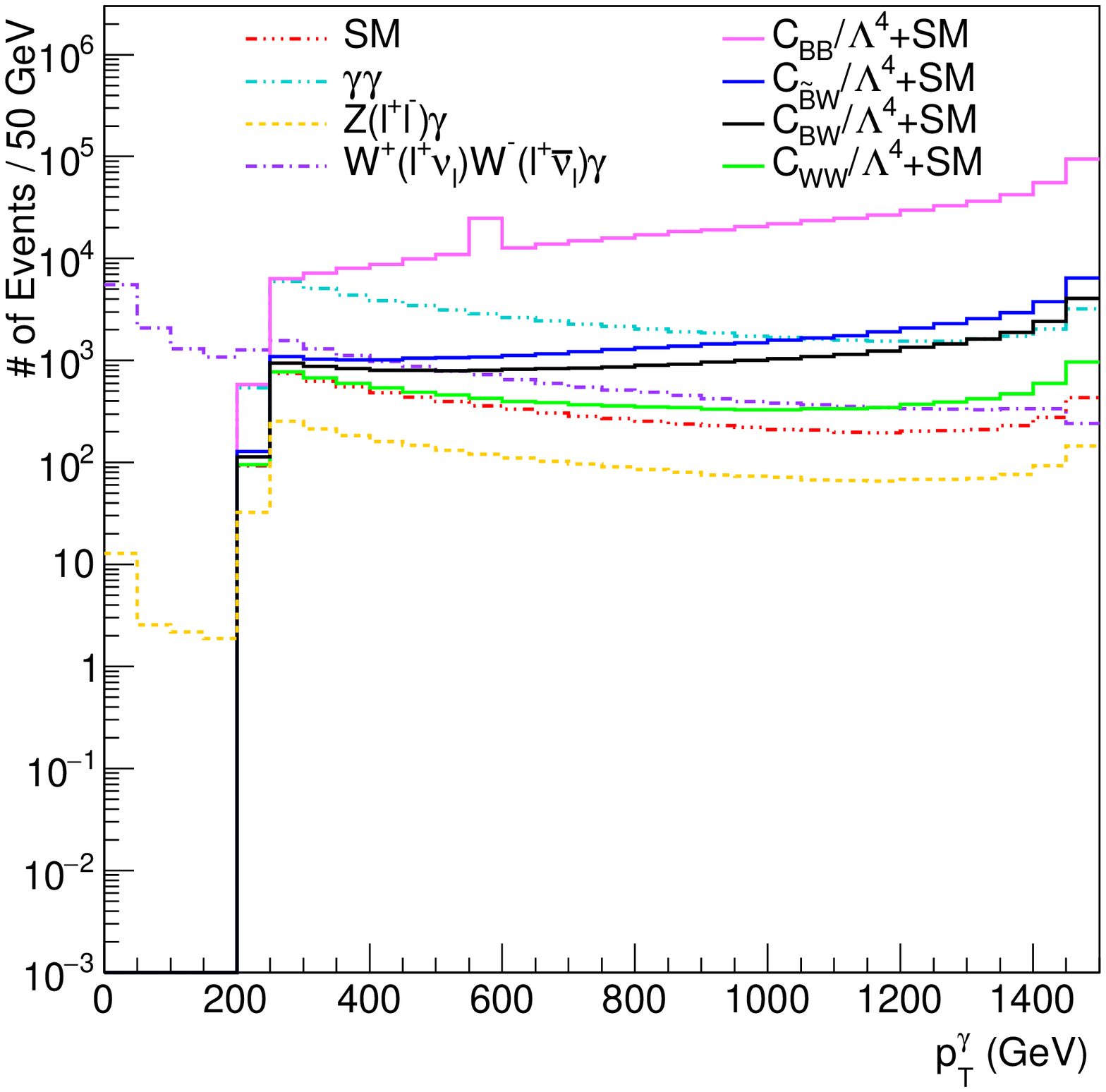}}}
\caption{The number of expected events (normalized to the $L_{int}$=1 ab$^{-1}$) as a function of $p^{\gamma}_T$ for the $\mu^+\mu^- \to Z\gamma$ signal and backgrounds. Here, $C_{\widetilde{B}W}/{\Lambda^4}$, $C_{BB}/{\Lambda^4}$, $C_{BW}/{\Lambda^4}$ and $C_{WW}/{\Lambda^4}$ couplings are equal to 1 TeV$^{-4}$. Here, the dotted lines represent the background processes, while the solid lines indicate the signal processes.}
\label{Fig.3}
\end{figure}

\begin{figure}[ht]
\centerline{\scalebox{1.4}{\includegraphics{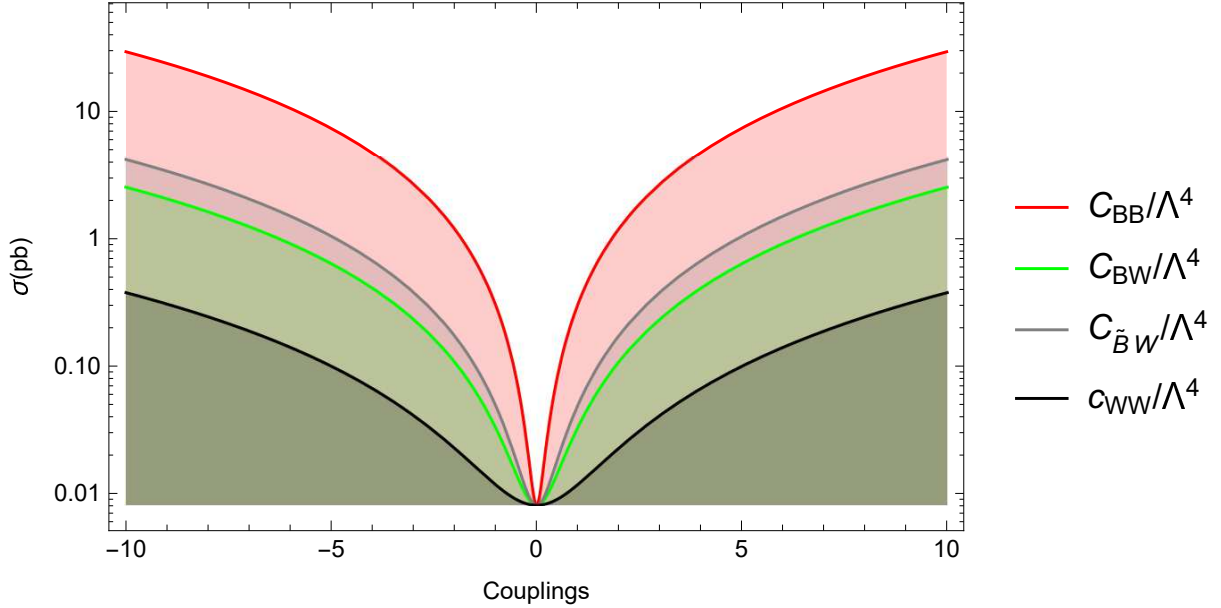}}}
\caption{The total cross-sections for the process $\mu^+\mu^-\to Z\gamma$ in terms of the anomalous $C_{BB}/\Lambda^{4}$, $C_{BW}/\Lambda^{4}$, $C_{\widetilde{B}W}/\Lambda^{4}$, $C_{WW}/\Lambda^{4}$ couplings at $\sqrt{s}=3$ TeV. Here, the values of coupling are in unit of TeV$^{-4}$.}
\label{Fig.4}
\end{figure}

The SM background includes contributions from the last two Feynman diagrams in Fig.~\ref{Fig.1}. The $\gamma \gamma$ is considered as a background candidate if one of these photons cannot be detected by the detector. The $\ell\ell\gamma$ is assumed to be another relevant background if both leptons cannot be detected by the detector. Finally, the $WW\gamma$ is conceived as background because of having a dilepton and a neutrino pair with the leptonic decay of $W$ bosons. However, we apply a set of kinematic cuts for the photon and missing energy in the final state of the examined process to distinguish the signal from the backgrounds mentioned above: $\eta^{\gamma}$, $p^\gamma_T$ and $\slashed{E}_T$ where $\eta^{\gamma}$ and $p^\gamma_T$ are the pseudo-rapidity and the transverse momentum of the photon and $\slashed{E}_T$ is the missing energy transverse. The number of expected events as a function of $\slashed{E}_T$ and $p^\gamma_T$ for signal and relevant backgrounds of the process $\mu^-\mu^+\,\rightarrow\,Z\gamma \rightarrow\, \nu\bar{\nu}\gamma$ at the muon collider with $\sqrt{s}$=3 TeV with $L_{int}$=1 ab$^{-1}$ are given in Figs.~\ref{Fig.2} and ~\ref{Fig.3}, respectively. Here, $C_{\widetilde{B}W}/{\Lambda^4}$, $C_{BB}/{\Lambda^4}$, $C_{BW}/{\Lambda^4}$ and $C_{WW}/{\Lambda^4}$ couplings are equal to 1 TeV$^{-4}$ as well as the dotted lines represent the background processes, while the solid lines indicate the signal processes. As can be seen from these figures, the signals are almost separated from the relevant backgrounds at values of $\slashed{E}_T>300$ GeV and $p^\gamma_T>300$ GeV. Thus, in this study, we use the kinematic cuts $|\eta^{\gamma}|< 2.5$, $\slashed{E}_T>300$ GeV and $p^\gamma_T>300$ GeV to obtain the sensitivities on each anomalous coupling. Also, the total cross-sections of the process $\mu^-\mu^+\,\rightarrow\,Z\gamma$ by using above mentioned cuts as a function of the anomalous $C_{\widetilde{B}W}/{\Lambda^4}$, $C_{BB}/{\Lambda^4}$, $C_{BW}/{\Lambda^4}$ and $C_{WW}/{\Lambda^4}$ couplings are given in Fig.~\ref{Fig.4}. Here, while only one of the aNTGC is non-zero at any time, the other is fixed to zero. For example, $C_{BB}/\Lambda^{4}$ is not equal to zero in the calculations, while  $C_{BW}/\Lambda^{4}$, $C_{\widetilde{B}W}/\Lambda^{4}$, $C_{WW}/\Lambda^{4}$ couplings are zero. Similar calculations are made in the other couplings. As can be seen from Fig.~\ref{Fig.4}, the deviation from the SM of the anomalous $C_{BB}/{\Lambda^4}$ coupling is larger than the others.

The sensitivities at 95$\%$ C. L.  on $C_{\widetilde{B}W}/{\Lambda^4}$, $C_{BB}/{\Lambda^4}$, $C_{BW}/{\Lambda^4}$ and $C_{WW}/{\Lambda^4}$ couplings defining the aNTGC $ZZ\gamma$ and $Z\gamma\gamma$ through the process $\mu^-\mu^+\,\rightarrow\,Z\gamma$ are obtained by $\chi^2$ analysis;

\begin{eqnarray}
\label{eq.27}
\chi^{2}=\sum_{i}^{n_{bins}} (\frac{N_{i}^{NP}-N_{i}^{B}}{N_{i}^{B}\Delta_{i}})^{2}
\end{eqnarray}

{\raggedright where $N_{i}^{NP}$ is the total number of events  of the anomalous couplings, $N_{i}^{B}$ is total number of events of the SM background. On the other hand, $\Delta_{i}=\sqrt{\delta_{sys}^{2}+\frac{1}{N_{i}^{B}}}$
is the combined systematic ($\delta_{sys}$)
and statistical uncertainties in each bin.}

The source of systematic uncertainties are mainly based on the cross section measurements of signal and background processes with leading-order (LO) or next to leading order (NLO) predictions and higher order electroweak (EW) corrections, the uncertainty in integrated luminosity, photon identification efficiency, as well as the uncertainties in the energy momentum scales and resolutions of the final-state particles. Discussing the sources of systematic uncertainty in detail is not one of the main aims of this study, but to investigate the overall effects of the systematic uncertainty on the limits values of aNTGC, we consider two different scenarios of systematic uncertainty. The systematic uncertainty given at the LHC with $\sqrt{s}$=13 TeV with $L_{int}$=36.1 fb$^{-1}$ while investigating the anomalous $ZZ\gamma$ and $Z\gamma\gamma$ couplings through the process $pp\,\rightarrow\,Z \gamma \rightarrow\, \nu \bar{\nu} \gamma$ is between 3.5$\%$ and 4.2$\%$ \cite{Aaboud:2018ybz}. For this reason, we consider that the systematic uncertainty for the process $\mu^-\mu^+\,\rightarrow\,Z\gamma \rightarrow\, \nu \bar{\nu} \gamma$ varies from $\delta_{sys}=3\%$ to 5$\%$.

\begin{figure}[h!]
\includegraphics[scale=0.4]{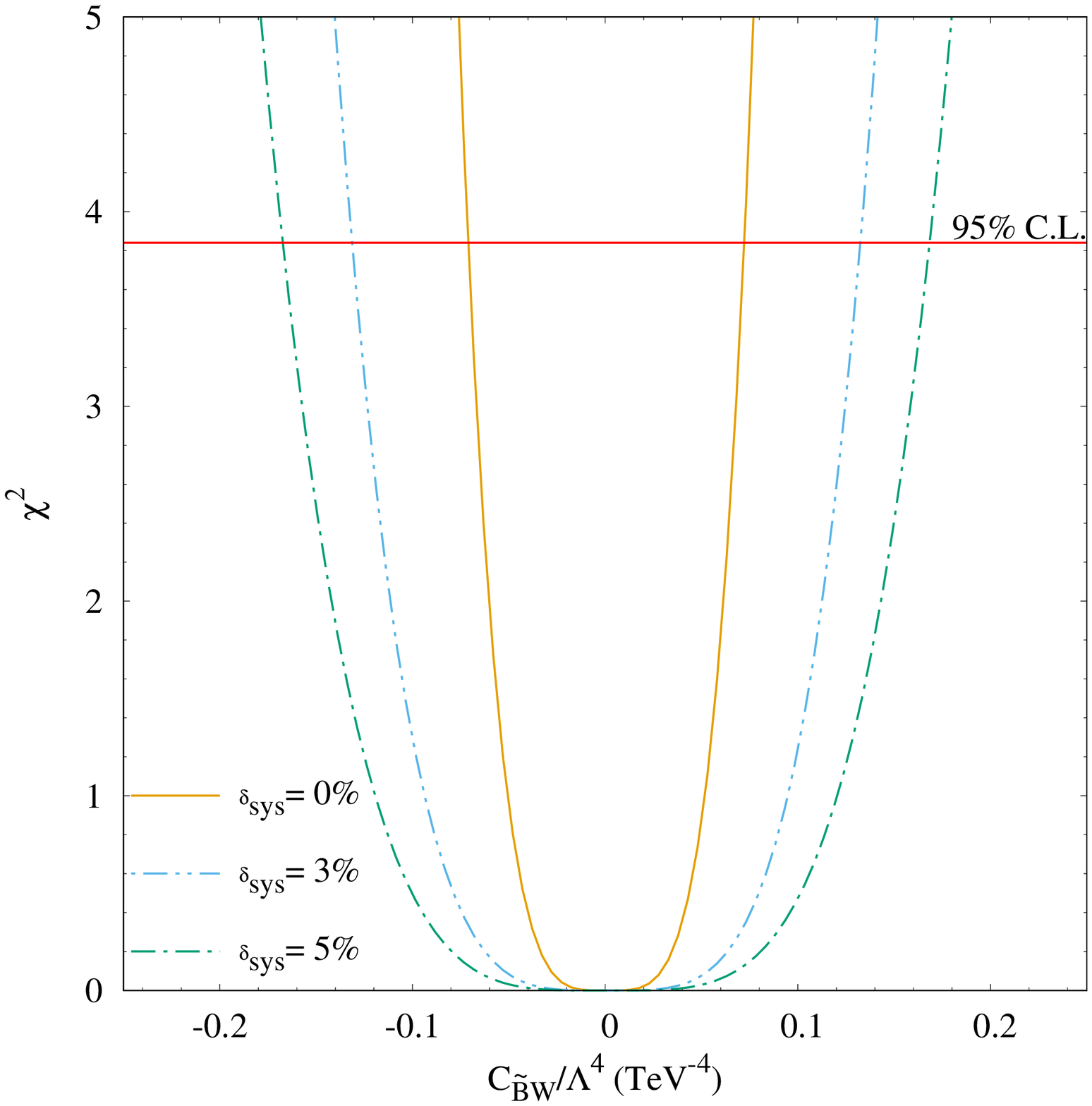}\includegraphics[scale=0.4]{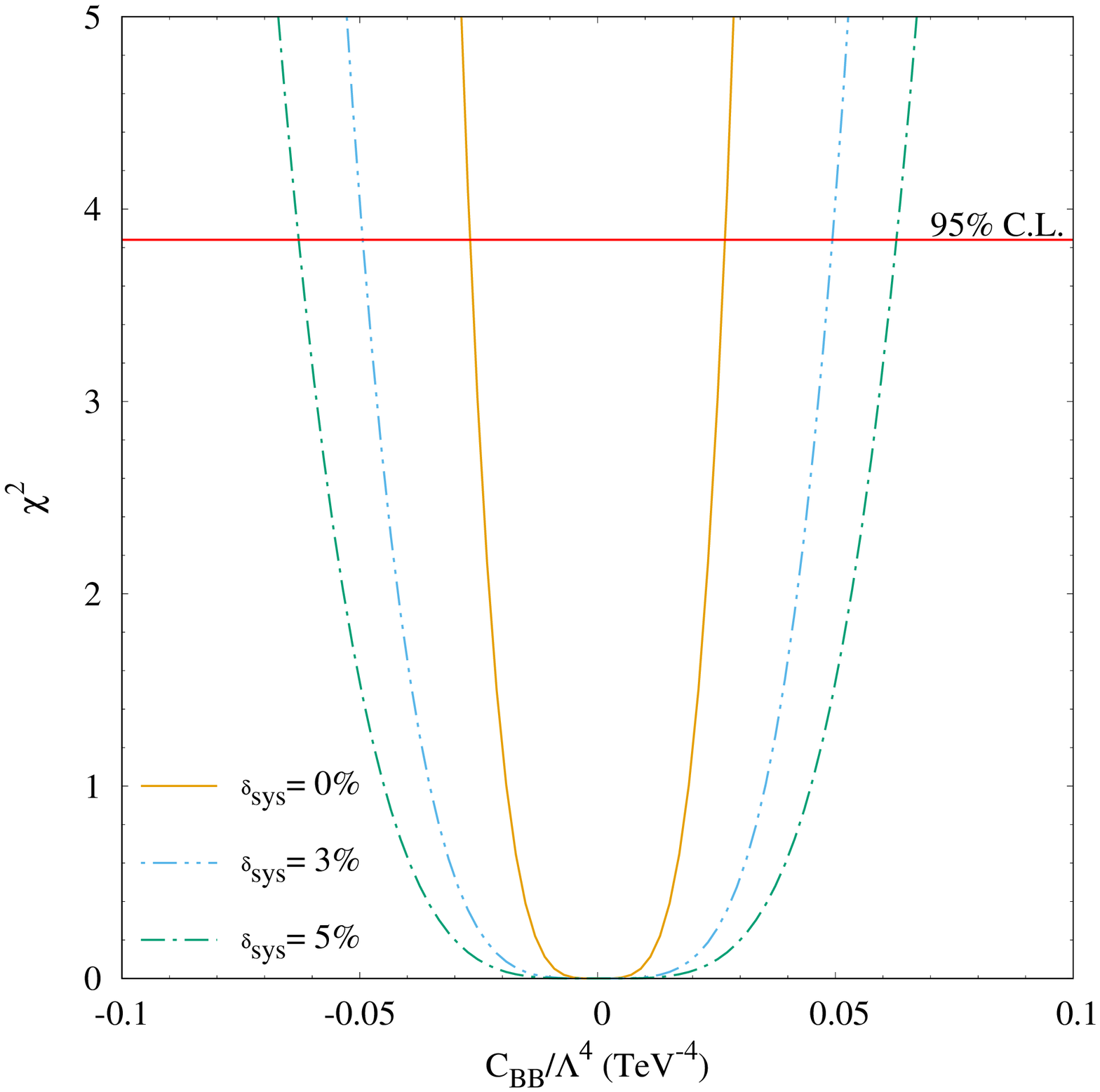}\\
\includegraphics[scale=0.4]{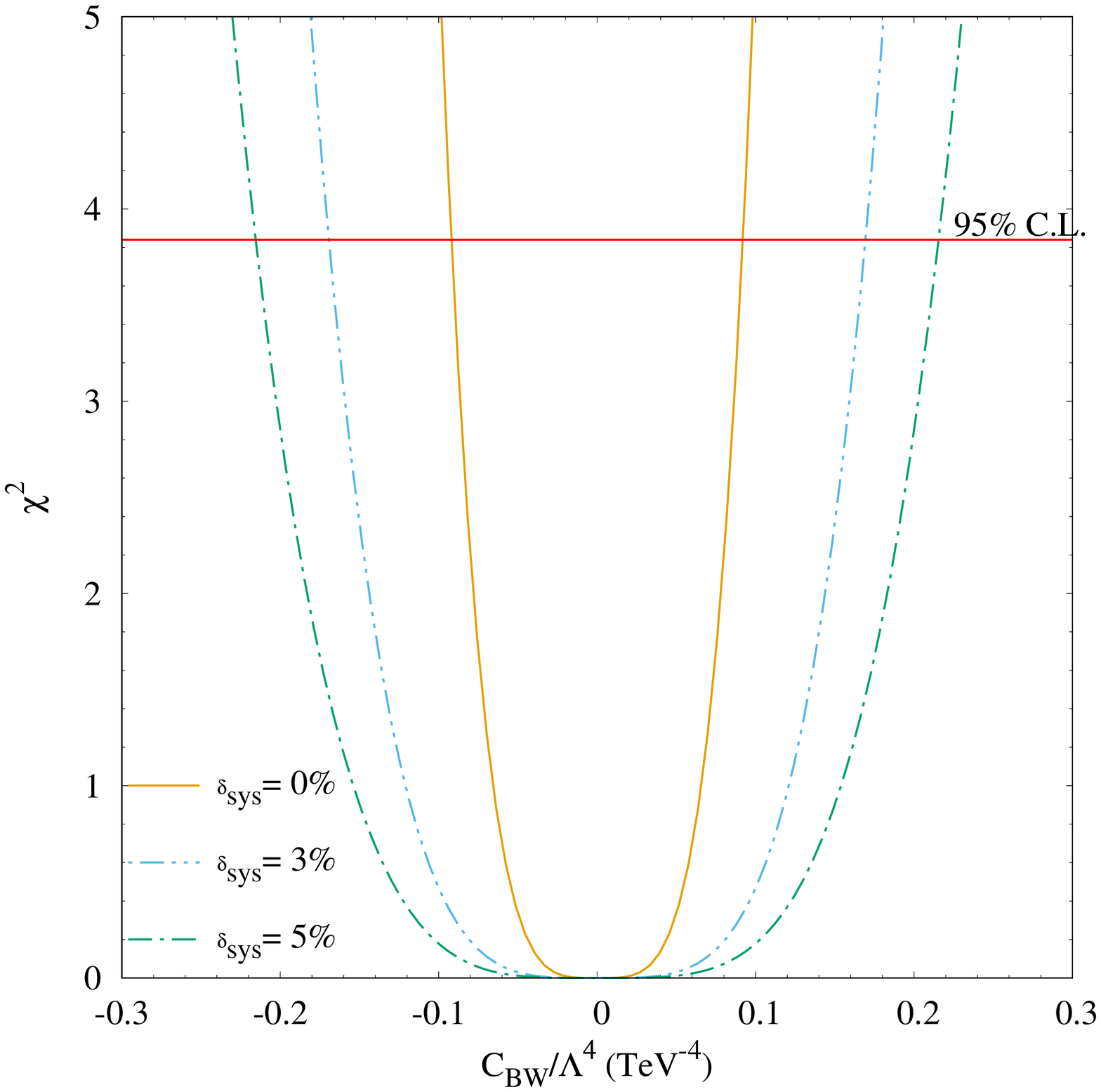}\includegraphics[scale=0.4]{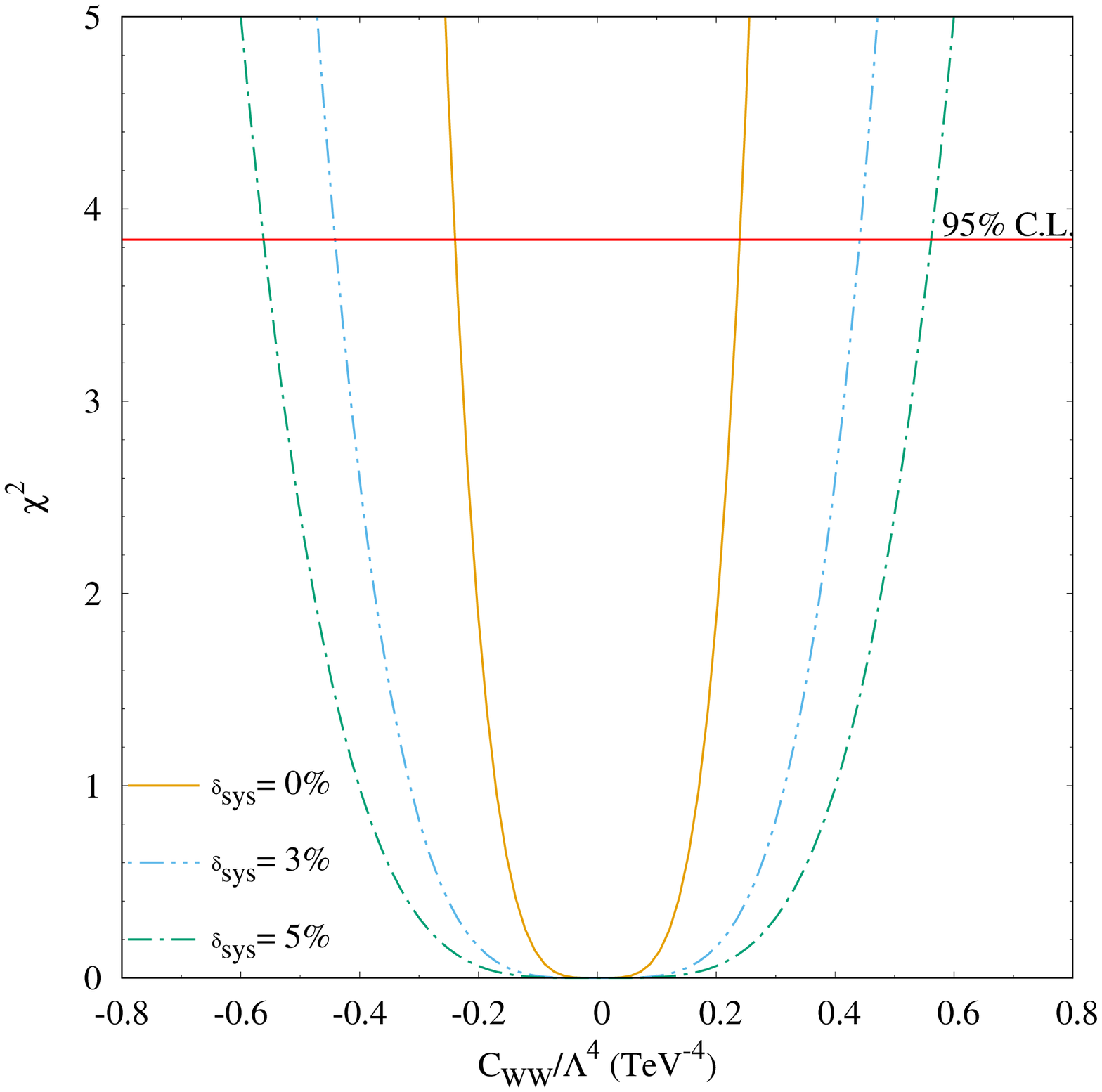}
\caption{$\chi^{2}$ values in terms of the anomalous $C_{\widetilde{B}W}/\Lambda^{4}$, $C_{BB}/{\Lambda^4}$, $C_{BW}/{\Lambda^4}$, $C_{WW}/{\Lambda^4}$ couplings under various systematic uncertainties.}
\label{Fig.5}
\end{figure}

Figs.~\ref{Fig.5} represents obtained $\chi^{2}$ values as a function of $C_{\widetilde{B}W}/{\Lambda^4}$, $C_{BB}/{\Lambda^4}$, $C_{BW}/{\Lambda^4}$ and $C_{WW}/{\Lambda^4}$ at the muon collider with $\sqrt{s}$=3 TeV with $L_{int}$=1 ab$^{-1}$ without and with systematic errors (3$\%$ and 5$\%$), respectively.

\begin{figure}[h!]
\centerline{\scalebox{0.75}{\includegraphics{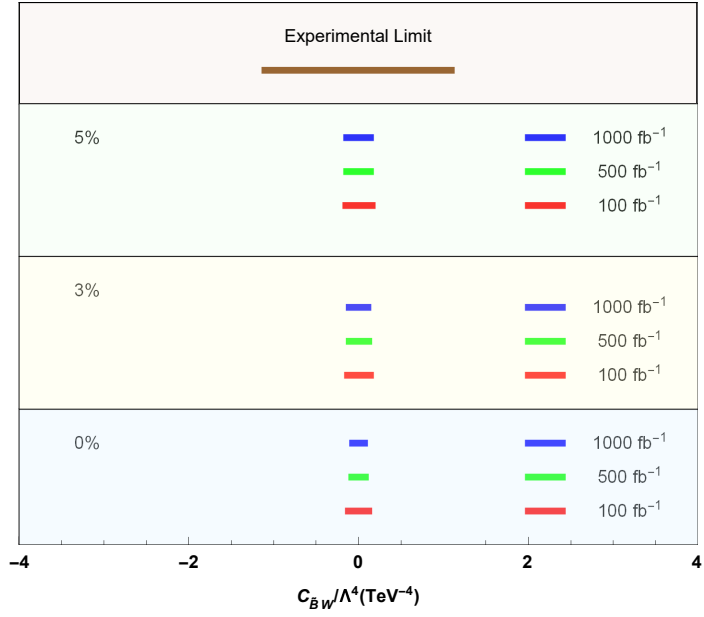}}}
\caption{Comparison of the ATLAS experimental limit and projected sensitivity on the anomalous $C_{\widetilde{B}W}/\Lambda^4$ coupling according to three different luminosities and systematic uncertainties.}
\label{Fig.9}
\end{figure}

\begin{figure}[h!]
\centerline{\scalebox{0.75}{\includegraphics{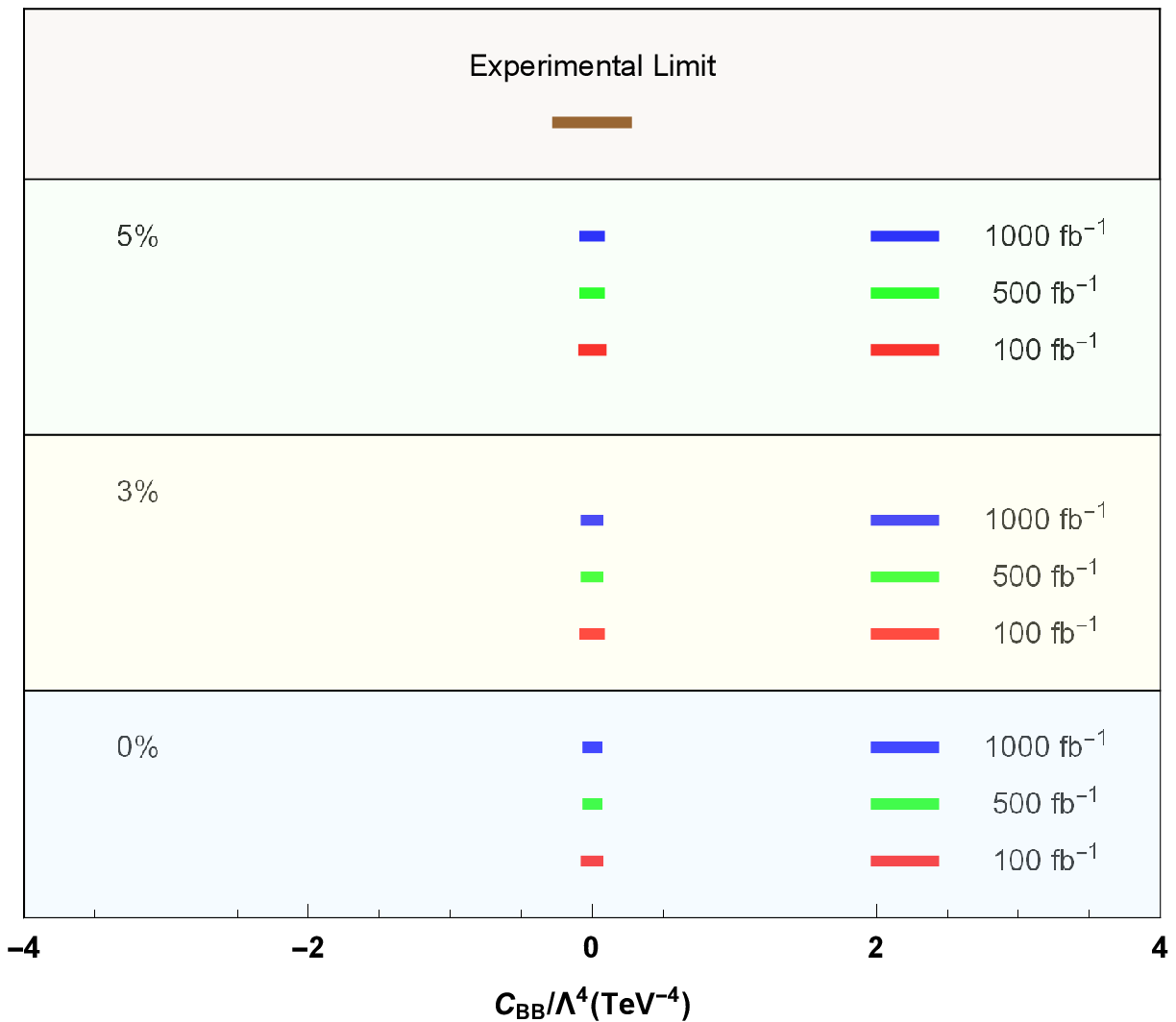}}}
\caption{Same as Fig.~\ref{Fig.9} but for the anomalous $C_{BB}/\Lambda^4$} coupling.
\label{Fig.10}
\end{figure}

\begin{figure}[h!]
\centerline{\scalebox{0.75}{\includegraphics{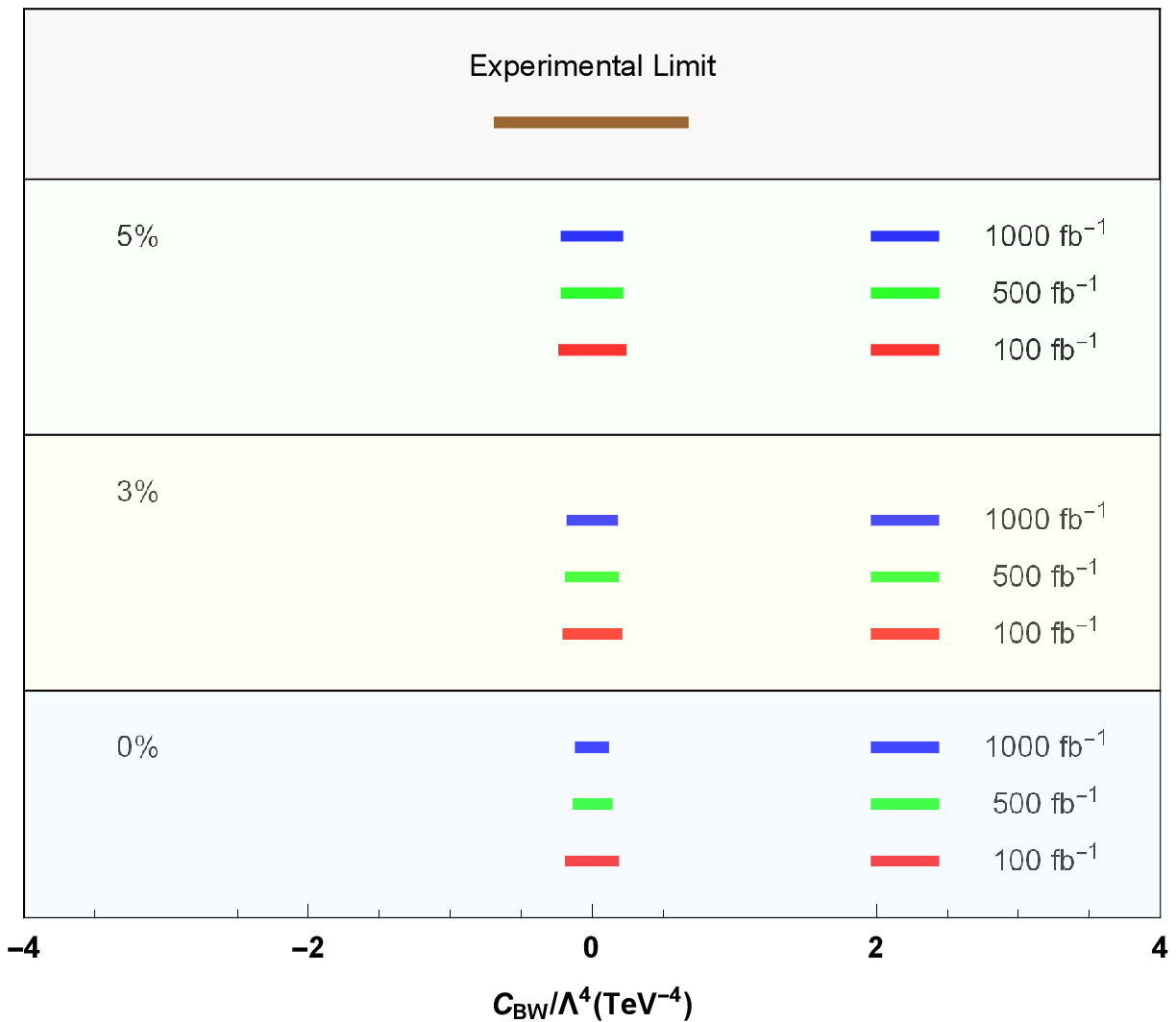}}}
\caption{Same as Fig.~\ref{Fig.9} but for the anomalous $C_{BW}/\Lambda^4$} coupling.
\label{Fig.11}
\end{figure}

\begin{figure}[h!]
\centerline{\scalebox{0.75}{\includegraphics{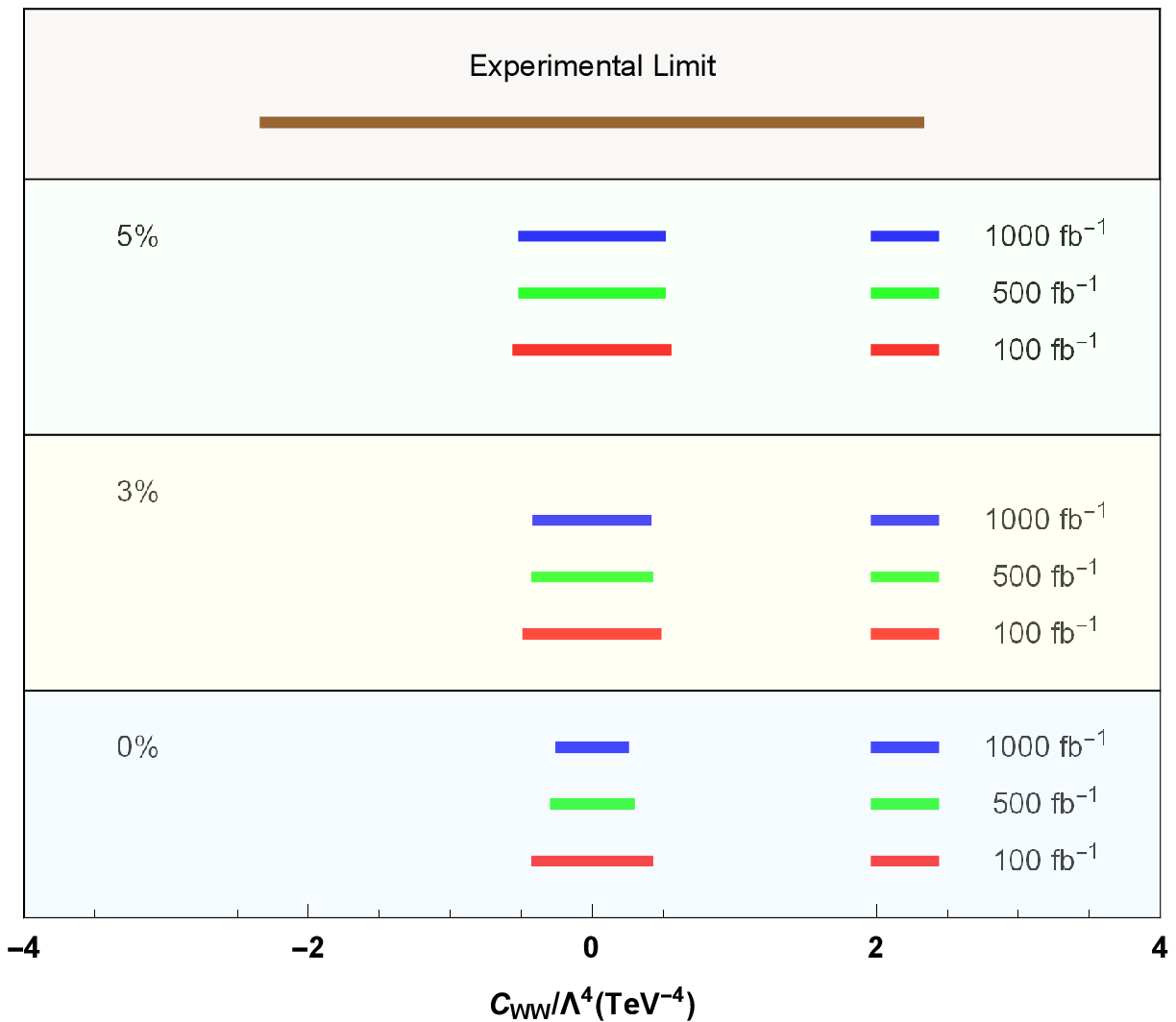}}}
\caption{Same as Fig.~\ref{Fig.9} but for the anomalous $C_{WW}/\Lambda^4$} coupling.
\label{Fig.12}
\end{figure}

Our best sensitivities obtained 95$\%$ C.L. on the aNTGC $C_{\widetilde{B}W}/{\Lambda^4}$, $C_{BB}/{\Lambda^4}$, $C_{BW}/{\Lambda^4}$ and $C_{WW}/{\Lambda^4}$ are $[-6.53;6.64]\times10^{-2}$ TeV$^{-4}$, $[-2.47;2.47]\times10^{-2}$ TeV$^{-4}$, $[-8.46;8.46]\times10^{-2}$ TeV$^{-4}$ and $[-2.20;2.20]\times10^{-1}$ TeV$^{-4}$, respectively. If the limits obtained on all anomalous couplings are compared, as we expect, the best sensitivity belongs to the anomalous $C_{BB}/{\Lambda^4}$ coupling. Also, Figs. 6-9 show the comparison of the current experimental limits and the sensitivities of the aNTGC $C_{\widetilde{B}W}/{\Lambda^4}$, $C_{BB}/{\Lambda^4}$, $C_{BW}/{\Lambda^4}$ and $C_{WW}/{\Lambda^4}$ according to three different luminosities and systematic uncertainties.  As can be seen from these figures the sensitivity of the anomalous couplings gradually improves as the integrated luminosity increases and the systematic uncertainties decrease. For example, the sensitivities on $C_{\widetilde{B}W}/\Lambda^{4}$ with $\delta_{sys}=3\%$ for $L_{int}$=100, 500, 1000 fb$^{-1}$ are obtained as $[-1.34;1.35]\times10^{-1}$ TeV$^{-4}$,  $[-1.15;1.16]\times10^{-1}$ TeV$^{-4}$ and $[-1.12;1.13]\times10^{-1}$ TeV$^{-4}$, respectively. Also, the sensitivities on $C_{\widetilde{B}W}/\Lambda^{4}$ with $L_{int}$=1000 fb$^{-1}$ for  $\delta_{sys}=0,3,5 \%$ are found as $[-6.53;6.64]\times10^{-2}$ TeV$^{-4}$,  $[-1.12;1.13]\times10^{-1}$ TeV$^{-4}$ and $[-1.41;1.43]\times10^{-1}$ TeV$^{-4}$, respectively. 
However, the limits with increasing the luminosity on the anomalous
couplings do not increase proportionately to the luminosity due to the systematic error considered here. The reason for this situation is the systematic error (especially for $5 \%$) which is much bigger than the statistical error.
As a result, it is easily understood that future muon collider with $\sqrt{s}$=3 TeV leads to a remarkable improvement in the existing experimental limits on the aNTGC. Even considering systematic uncertainty of 5$\%$, the obtained limits for the muon collider are better than the LHC results on all couplings.

\section{Conclusions}

In this study, we present phenomenological cut-based research for investigating the limits on the $CP$-conserving $C_{\widetilde{B}W}/{\Lambda^4}$ coupling and $CP$-violating $C_{BB}/{\Lambda^4}$, $C_{BW}/{\Lambda^4}$, $C_{WW}/{\Lambda^4}$ couplings through $Z\gamma \to (\nu \nu) \gamma $ production at the muon collider. We show that $\slashed{E}_T$ and $p^\gamma_T$ are important to distinguish signal and relevant backgrounds. Applying the cut-based analysis, the sensitivities of each anomalous coupling are obtained by using $\chi^2$ method for the muon collider with $\sqrt{s}$=3 TeV with $L_{int}$=1 ab$^{-1}$. Comparing our results with experimental research for the production of $\nu\nu\gamma$ from LHC with $\sqrt{s}$=13 TeV, as well as phenomenological studies for the production of $\nu\nu\gamma$ and $4\ell$ from the HL-LHC, the HE-LHC, the FCC-hh, and the CLIC. While we have better results for all anomalous couplings obtained than the HL-LHC, the HE-LHC, and the FCC-hh, the sensitivities on the anomalous couplings are almost 2 times worse than the sensitivities obtained from the CLIC. As a result, the findings of this study show that the future muon collider with $\sqrt{s}$=3 TeV can improve the sensitivity limits on $C_{\widetilde{B}W}/{\Lambda^4}$, $C_{BB}/{\Lambda^4}$, $C_{BW}/{\Lambda^4}$ and $C_{WW}/{\Lambda^4}$ parameters defining the aNTGC $ZZ\gamma$ and $Z\gamma\gamma$ with respect to the experimental limits of LHC and the phenomenological limits of future hadron-hadron colliders.
\newline

\textbf{Data Availability Statement} The datasets generated during and analysed during the current study are available from the corresponding author on reasonable request.

\end{document}